\documentclass[aps,prd,amsmath,amssymb,superscriptaddress,nofootinbib]{revtex4-2}

\usepackage{graphicx} 
\usepackage{dcolumn}  
\usepackage{bm}       
\usepackage{physics}  
\usepackage{hyperref} 
\usepackage{booktabs}
\begin{document}

\title{Distinguishing Dirac and Majorana Neutrinos: Resonant Spin-Flavor Precession of GeV-Scale Astrophysical Transients}

\author{David Delepine}
\email{delepine@ugto.mx}
\affiliation{Divisi\'on de Ciencias e Ingenier\'ias, Universidad de Guanajuato, C.P.
37150, Le\'on, Guanajuato, M\'exico.}

\author{A. Yebra}
\email{azarael@fisica.ugto.mx}
\affiliation{Divisi\'on de Ciencias e Ingenier\'ias, Universidad de Guanajuato, C.P.
37150, Le\'on, Guanajuato, M\'exico.}

\date{\today}

\begin{abstract}
We present a unified formalism to study the Resonant Spin-Flavor Precession (RSFP) of high-energy ($\sim 1$ GeV) transient astrophysical neutrinos as a probe of their fundamental Dirac or Majorana nature.
Current MeV-scale neutrino studies face stringent restrictions: efficient core RSFP for Dirac supernova neutrinos is excluded by SN1987A cooling bounds, while solar neutrino conversion is tightly constrained by Borexino data.
We show that considering the 1 GeV energy scale—accessible through solar flares modifies the resonance conditions.
For 1 GeV solar flare neutrinos, the resonance shifts to the tachocline and convective zones, where toroidal magnetic fields ($B \sim 50$ kG) induce adiabatic spin-flavor conversion.
In contrast, for supernovae, to avoid the cooling constraints, the RSFP is moved from the core supernovae to the stellar envelope.
As for non-thermal 1 GeV supernova neutrinos, the resonance is located in the dilute stellar wind where magnetic fields are negligible, suppressing RSFP and preserving the flux, one can use these non-thermal neutrinos as a candle to calibrate our signal, reducing its dependence on astrophysical uncertainties.
Evaluating these helicity transitions through a density matrix approach, we predict distinct asymmetries in Coherent Elastic Neutrino-Nucleus Scattering (CE$\nu$NS) and neutrino-electron scattering cross-sections for solar flare neutrinos and supernova neutrinos.
Our proposal provides a viable method to distinguish Dirac from Majorana neutrinos and to probe magnetic moments down to $\mu_\nu \sim 10^{-14} \mu_B$.
\end{abstract}
\maketitle

\section{Introduction}

The observation of neutrino oscillations has established that neutrinos possess mass \cite{Super-Kamiokande:1998kpq, KamLAND:2002uet, SNO:2002tue}.
However, determining whether they are Dirac fermions \cite{Dirac:1928hu, Weyl:1929fm, Pauli:1930pc} or Majorana fermions \cite{Majorana:1937vz, Bilenky:1987ty} remains an open question in the Standard Model \cite{Giunti:2014ixa}.
Neutrinoless double beta decay ($0\nu\beta\beta$) experiments \cite{Racah:1937qq, Furry:1939qr, Pontecorvo:1957qd, Schechter:1981bd} are the standard avenue for investigating this, though their sensitivity depends on the absolute neutrino mass scale, which might be quite small.Alternative proposals include using CE$\nu$NS at reactor or accelerator experiments \cite{Hati:2022CEvNS}.

An alternative diagnostic approach relies on the electromagnetic properties of the neutrino \cite{Fujikawa:1980yx, Shrock:1982sc, Schechter:1981hw, Pal:1981rm}.
If neutrinos have a non-zero magnetic moment $\mu_\nu$, their interaction with intense astrophysical magnetic fields can induce Resonant Spin-Flavor Precession (RSFP) \cite{Cisneros:1970nq, Okun:1986hi, Lim:1987tk, Akhmedov:1988uk, Balantekin:1990jb}.
Such a mechanism can rotate the spin of an active left-handed neutrino into a right-handed state.
The phenomenological consequences depend on the particle type: Dirac neutrinos convert into sterile states, whereas Majorana neutrinos convert into active right-handed antineutrinos that can still interact in terrestrial detectors.
The use of RSFP as a probe has typically been limited by constraints in standard thermal ($E \sim 10$ MeV) energy regimes.
While the application of RSFP to core-collapse supernovae has been explored \cite{Raffelt:1999tx, Akhmedov:2002mf, Ando:2003ie}, efficient core conversion of Dirac neutrinos into sterile states would lead to a rapid energy drain.
This conflicts with the observed 10-second diffusion timescale from SN1987A, placing bounds on the neutrino magnetic moment \cite{Nussinov:1987pc, Lattimer:1988mf, Barbieri:1988av, Kuznetsov:2009jz, Raffelt:2001kv, Burrows:1988ah}.
Furthermore, the non-observation of solar antineutrinos by Borexino restricts significant spin-flavor conversion for standard $^8$B neutrinos \cite{Borexino:2008gab, Montanino:2008par, Borexino:2017fbd}.

In the present work, we extend the RSFP formalism to the case of high-energy ($\sim 1$ GeV) transient neutrinos, such as those produced in solar flares and the non-thermal tails of core-collapse supernovae.
As the density required for matter-induced resonance scales inversely with the neutrino energy ($\rho_{\text{res}} \propto \Delta m^2 / E_\nu$), the 1 GeV scale shifts the resonance zone in these astrophysical environments:
\begin{itemize}
    \item \textbf{Solar Environment:} For 1 GeV flare neutrinos, the resonance is relocated from the dense radiative core to the tachocline and convective zones, where the solar dynamo produces significant magnetic fields ($B \sim 10^3 - 10^5$ G) \cite{Spiegel:1992, Basu:1997zz}.
\item \textbf{Supernova Environment:}  To avoid the cooling supernovae constraints, the RSFP is moved from the core supernovae to the stellar envelope.
The resonance for the 1 GeV non-thermal tail \cite{Murase:2018, Wen:2024} moves out of the dense envelope into the extended stellar wind.
The negligible magnetic fields in this region \cite{Jackson:1998} lead to an adiabaticity drop such that the neutrino spin state is frozen.
So the 1 GeV non-thermal supernovae neutrinos are used to calibrate the signal, reducing the dependence of the experimental results on astrophysical uncertainties.
\end{itemize}

We analyze the propagation of these high-energy neutrinos by treating the ensemble as an open quantum system \cite{Burgess:1996mz, Balantekin:2004tk, Studenikin:2018vnp} and adopting a density matrix formalism \cite{McKellar:1992ja, Stodolsky:1986dx, Sigl:1992fn}.
This formalism allows us to incorporate collisional decoherence, magnetohydrodynamic turbulence, and non-linear neutrino self-interactions in a unified way.
Within this setup, we compute the evolution of the helicity polarization and identify conditions under which RSFP can lead to sizeable spin-flavor conversion in the solar tachocline while leaving the GeV supernova tail essentially unaffected.
It is shown that this high-energy RSFP induces measurable asymmetries in scattering cross-sections, specifically in Coherent Elastic Neutrino-Nucleus Scattering (CE$\nu$NS).
In particular, we show that Coherent Elastic Neutrino-Nucleus Scattering (CE$\nu$NS) and neutrino–electron scattering at GeV energies can exhibit characteristic asymmetries once RSFP-induced helicity flips are taken into account.
By employing the non-resonant high-energy supernova tail as a standard candle, we discuss an experimental approach aimed at isolating the magnetic moment signature from astrophysical uncertainties.

The paper is organized as follows. In Sec.~\ref{sec:theory} we summarize the density matrix formalism and the effective Hamiltonian governing neutrino evolution in dense magnetized media.
In section~\ref{sec:solar_application} this formalism is applied to 1 GeV solar flare neutrinos, focusing on the role of different solar magnetic field profiles and on the unified helicity evolution through the tachocline and convective zone.
In Sec.~\ref{sec:supernova_application} we turn to core-collapse supernovae, contrasting the behavior of the thermal neutrino burst with that of the high-energy non-thermal tail. In Sec.~\ref{sec:event_rates} we provide order-of-magnitude event rate estimates and the prospects for detection at upcoming facilities are discussed.
In section~\ref{sec:scattering} the implications for neutrino–electron scattering and CE$\nu$NS are discussed, and a ratio observable designed to distinguish Dirac from Majorana neutrinos is introduced.
We conclude in Sec.~\ref{sec:conclusion} with a summary and an outlook on the parameter space that could be probed by this strategy.
\section{Neutrino Evolution in Dense, Magnetized Media}
\label{sec:theory}

To describe neutrino propagation in realistic astrophysical settings, it is convenient to treat the ensemble as an open quantum system \cite{Burgess:1996mz, Balantekin:2004tk, Studenikin:2018vnp}.
The usual Schr\"odinger wave-function approach assumes perfect coherence and is therefore inadequate once one includes the combined effects of large matter densities, magnetic turbulence, and collective neutrino self-interactions. In what follows we work within the density matrix formalism \cite{McKellar:1992ja, Stodolsky:1986dx, Sigl:1992fn}. 

\subsection{Lindblad Master Equation and Effective Hamiltonian}

The time evolution of the density matrix $\rho_{\mathbf{p}}$ for a momentum mode $\mathbf{p}$ is governed by a Lindblad-type master equation:
\begin{equation}
i \frac{d\rho_{\mathbf{p}}}{dt} = [H_{\text{eff}}, \rho_{\mathbf{p}}] - i\mathcal{D}[\rho_{\mathbf{p}}],
\label{eq:lindblad}
\end{equation}
where $\mathcal{D}[\rho_{\mathbf{p}}]$ denotes the dissipator functional that encodes collisional decoherence and depolarization induced by magnetohydrodynamic (MHD) turbulence \cite{Lisi:2000zt, Friedland:2000rn, Oliveira:2019hyn, Kurashvili:2017zpz}. The first term describes coherent evolution under an 
effective Hamiltonian $H_{\text{eff}}$ and the second term accounts for environmental effects that drive the system away from pure states.
The effective Hamiltonian is given as a sum of four contributions,
\begin{equation}
H_{\text{eff}} = H_{\text{vac}} + H_{\text{mat}} + H_{\text{mag}} + H_{\nu\nu},
\label{eq:hamiltonian_full}
\end{equation}
where $H_{\text{vac}}$ is the vacuum oscillation term, $H_{\text{mat}}$ contains the matter-induced potentials, $H_{\text{mag}}$ encodes magnetic moment interactions, and $H_{\nu\nu}$ represents non-linear neutrino self-interactions which may further modify the evolution in the dense supernova environment \cite{Duan:2010bg, Hannestad:2006nj, Banerjee:2011fj, Mirizzi:2021MagMomCollective}. While our density matrix formalism is built to accommodate non-linear self-interactions, for the purposes of this exploratory numerical study, we neglect $H_{\nu\nu}$ to focus on the interplay between resonance shifts and magnetic profiles.
For our purposes, it is convenient to work in a two-state flavor–spin basis $(\nu_{eL}, \nu_{xR})^T$, where $\nu_{xR}$ denotes a generic right-handed state.

For our numerical analysis, we focus on the evolution in a two-state flavor-spin basis $(\nu_{eL}, \nu_{xR})^T$, where $\nu_{xR}$ represents a generic right-handed state[cite: 60]. In this  basis, the sub-Hamiltonian  responsible for Spin-Flavor Precession (SFP) is:
\begin{equation}
H_{SFP}=\begin{pmatrix}V_{e}-\frac{\Delta m^{2}}{4E}\cos 2\theta&\mu_{\nu}B_{\perp}(r)\\ \mu_{\nu}B_{\perp}(r)&V_{x}+\frac{\Delta m^{2}}{4E}\cos 2\theta\end{pmatrix}
\end{equation}
Note that $H_{SFP}$ represents the coherent part of the evolution omitting the non-linear self-interaction term $H_{\nu\nu}$ introduced in Eq. (2), as collective effects are beyond the scope of this initial exploratory study.

In this basis, the part of the Hamiltonian responsible for Spin-Flavor Precession (SFP) \cite{Balantekin:2004tk, Sasaki:2023tgo} can be written as
\begin{equation}
H_{\text{SFP}} =
\begin{pmatrix}
V_e - \dfrac{\Delta m^2}{4E} \cos 2\theta & \mu_\nu B_\perp(r) \\
\mu_\nu B_\perp(r) & V_x + \dfrac{\Delta m^2}{4E} \cos 2\theta
\end{pmatrix}.
\end{equation}
Here $E$ is the neutrino energy, $\Delta m^2$ and $\theta$ are the atmospheric or solar mass-squared difference and mixing angle (depending on the channel), and $B_\perp(r)$ is the component of the magnetic field transverse to the direction of neutrino propagation.
The effective matter potential for electron neutrinos is
\begin{equation}
V_e = \sqrt{2} G_F \left(N_e - \frac{N_n}{2}\right),
\end{equation}
where $N_e$ and $N_n$ are the electron and neutron number densities, respectively. The matter potential $V_x$ is defined as:

\begin{align}
V_x &=
\begin{cases}
0 & \text{Dirac case (sterile right-handed state } \nu_R\text{)}, \\
0 & \text{Majorana case (active antineutrino state } \bar{\nu}_{\mu/\tau}\text{, up to small NC terms)}.
\end{cases}
\end{align}
In both cases we neglect subleading neutral-current differences, so that the main qualitative distinction comes from whether the right-handed state is sterile or remains active.
A detailed treatment of collective oscillations with non-zero magnetic moments has been carried out in Ref.~\cite{Mirizzi:2021MagMomCollective}, which is beyond the scope of our simplified treatment.

The core physical distinction lies in the nature of the right-handed state. In the Dirac scenario, $\nu_{xR}$ is a sterile singlet that does not participate in Standard Model interactions, leading to $V_x = 0$. In the Majorana scenario, the right-handed state is an active antineutrino ($\bar{\nu}_{\mu/\tau}$) which interacts via the Neutral Current (NC) potential, $V_{NC}$. 

However, in the astrophysical environments considered here—where electron density $N_e$ and neutron density $N_n$ are comparable—the effective potential difference is dominated by the Charged Current (CC) term $V_e = \sqrt{2} G_F (N_e - N_n/2)$. Since $V_{NC} \propto N_n$, its contribution to the resonance condition ($\Omega_3 \simeq 0$) is subleading. For this exploratory analysis, we take $V_x \approx 0$ for both cases to isolate the macroscopic signature: the total disappearance of flux in the Dirac case versus the preservation of active flux in the Majorana case.

\subsection{Spin-Flavor Polarization and Bloch Equation}

It is useful to parametrize the density matrix in terms of a polarization vector $\vec{P}(t)$ using the Pauli matrices $\vec{\sigma}$:
\begin{equation}
\rho(t) = \frac{1}{2} \left[ P_0(t) \mathbb{I} + \vec{P}(t) \cdot \vec{\sigma} \right].
\label{eq:rho_expansion}
\end{equation}
The component $P_0(t)$ measures the total occupation number, whereas the vector $\vec{P}(t)$ encodes the flavor–spin coherence.
For our analysis, the important quantity is the longitudinal helicity asymmetry:
\begin{equation}
S_{\parallel}(r) \equiv P_3(r),
\end{equation}
$S_{\parallel}(r)$ corresponds to the difference between right-handed and left-handed occupation probabilities, normalized to the total population.
Substituting Eq.~\eqref{eq:rho_expansion} into Eq.~\eqref{eq:lindblad} yields a generalized Bloch equation of the form
\begin{equation}
\frac{d\vec{P}}{dt} = \vec{\Omega} \times \vec{P} - \Gamma \vec{P}_\perp,
\label{eq:bloch_vector}
\end{equation}
where $\vec{\Omega}$ is an effective precession vector and $\vec{P}_\perp$ is the component of $\vec{P}$ transverse to $\vec{\Omega}$.
The first term describes coherent precession of the polarization vector around $\vec{\Omega}$, while the second term with rate $\Gamma$ summarizes environmental depolarization effects.
The relevant components of $\vec{\Omega}$ can be identified as
\begin{align}
\Omega_1 &= 2\mu_\nu B_\perp(r), \\
\Omega_3 &= V_{\text{eff}}(r) - \frac{\Delta m^2}{2E} \cos 2\theta,
\end{align}
where $V_{\text{eff}}(r)= V_e - V_x$ is the effective matter potential entering the SFP channel.
The transverse component $\Omega_1$ controls the Larmor torque that drives spin-flavor transitions, while $\Omega_3$ governs the approach to resonance.
The depolarization parameter $\Gamma$ describes the cumulative effect of collisions and MHD-induced turbulence on the off-diagonal elements of $\rho_{\mathbf{p}}$.
\subsection{Resonance Condition and Adiabaticity}

Efficient spin-flavor conversion requires that the evolution pass sufficiently close to the resonance condition
\begin{equation}
\Omega_3 \simeq 0,
\end{equation}
which corresponds to an exact cancellation between the matter potential and the vacuum term in the effective mixing angle \cite{Dighe:1999bi, Schirato:2002tg, Fogli:2003dw}.
The character of the transition is governed by the adiabaticity parameter $\gamma$, defined at the resonance point as
\begin{equation}
\gamma = \frac{\Omega_1^2}{|\dot{\Omega}_3|}
= \frac{(2\mu_\nu B_\perp)^2}{\left|\dfrac{d}{dr}(V_{\text{eff}})\right|_{\text{res}}}.
\end{equation}
Within the standard Landau–Zener approximation \cite{Landau:1932, Zener:1932ws, Parke:1986jy, Petcov:1987cd}, the non-adiabatic jump probability is
\begin{equation}
P_{LZ} = \exp\left(-\frac{\pi}{2} \gamma\right),
\end{equation}
and the final helicity asymmetry can be related to $\gamma$ through
\begin{equation}
S_{\parallel} = 1 - 2 \exp\left(-\frac{\pi}{2} \gamma\right).
\end{equation}
Large values of $\gamma$ correspond to adiabatic evolution and thus to efficient spin-flavor conversion, whereas small $\gamma$ imply that the system remains close to its initial state.
Since the resonance density scales approximately as $\rho_{\text{res}} \propto E^{-1}$, moving from MeV to GeV energies shifts the spatial location of the resonance in a given astrophysical environment.
As a consequence, the gradient $\left|\dfrac{d}{dr}(V_{\text{eff}})\right|_{\text{res}}$ and the local magnetic field $B_\perp(r)$ are sampled in very different regions, which can alter the adiabaticity parameter.
In the next sections we exploit this energy dependence to show how GeV-scale neutrinos can probe magnetized layers in the Sun and in how supernova neutrinos can have an efficient RSFP transition in the stellar envelope.
\section{Solar Flare Neutrinos and the Tachocline}
\label{sec:solar_application}

For standard solar neutrinos with energies $E \sim 10$ MeV, the RSFP resonance condition is typically satisfied deep inside the radiative core, at radii $r \lesssim 0.2 R_\odot$.
In that regime, Borexino data on the $^8$B flux strongly constrains any sizable active–sterile or active–antineutrino conversion \cite{Borexino:2017fbd}, leaving little room for RSFP to operate without conflicting with observations.
Transient solar flares provide a qualitatively different situation. 

\subsection{Solar Magnetic Field Profiles}

The RSFP transition probabilities depend sensitively on the spatial profile of the solar magnetic field $B(r)$.
So, we adopt three phenomenological models that are representative of different regions in the solar interior and convection zone:
\begin{itemize}
\item \textbf{Model I: Wood–Saxon profile (radiative zone).} In this case we assume a relic magnetic field confined mainly to the core, parameterized as
\begin{equation}
B(r) = \frac{B_0}{1 + \exp\left[k\left(\frac{r}{R_\odot} - R_{\text{cut}}\right)\right]},
\end{equation}
with $R_{\text{cut}} \approx 0.71 R_\odot$ and $k = 20$ \cite{Couvidat:2002, Friedland:2002ph}.
This profile captures a relatively sharp decline in the field strength near the base of the convective zone.
\item \textbf{Model II: Gaussian tachocline profile.} Here the magnetic field is associated with the solar dynamo operating in the shear layer around the tachocline.
It is modeled as
\begin{equation}
B(r) = B_{\text{max}} \exp\left[-\frac{(r - r_{\text{tach}})^2}{2\sigma^2}\right],
\end{equation}
with $r_{\text{tach}} \approx 0.71 R_\odot$ and peak field $B_{\text{max}} \sim 20$–$50$ kG \cite{Spiegel:1992, Basu:1997zz, Charbonneau:2010zz}.
The width $\sigma$ parametrizes the extent of the tachocline region over which strong toroidal fields are maintained.
\item \textbf{Model III: Convective-zone power law.} For radii $r > R_{\text{cut}}$, we describe the field in the convection zone and above by a decaying power law,
\begin{equation}
B(r) = B_s \left(\frac{R_\odot}{r}\right)^n,
\end{equation}
supplemented by stochastic turbulent fluctuations $\delta B$ \cite{Kitchatinov:1993, Fan:2009zz, Hathaway:2010zz}.
This model is intended to mimic the large-scale field in the convective envelope together with smaller-scale MHD turbulence.
\end{itemize}
Although these profiles are simplified, they are sufficient for our purpose of illustrating how the resonance location at GeV energies samples different magnetic environments than in the standard MeV regime.
\subsection{Solar Flare Flux and Resonance Shift}

\begin{figure}[ht]
\centering
\includegraphics[width=0.9\linewidth]{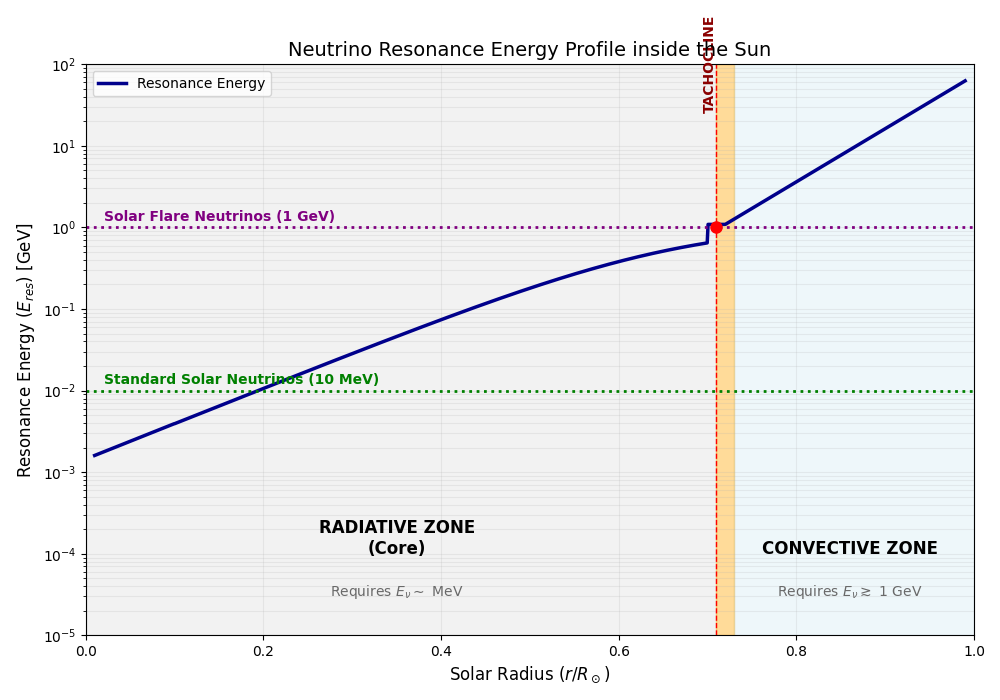}
\caption{Neutrino resonance energy ($E_{res}$) as a function of solar radius, calculated for standard LMA oscillation parameters ($\Delta m^2 \approx 7.5 \times 10^{-5} \text{ eV}^2$ ) and a solar mixing angle of $\theta_{12} \approx 34^{\circ}$. The horizontal dashed lines indicate the resonance regions for the standard $^8$B flux (10 MeV) and the transient solar flare flux (1 GeV). The visible discontinuity at $r/R_{\odot} \approx 0.71$ corresponds to the transition in the solar density profile at the base of the convective zone (the tachocline), where the abrupt change in the electron density gradient impacts the resonance condition calculation.
}
\label{fig:resonance_condition}
\end{figure}

Large solar flares accelerate protons and heavier ions via magnetic reconnection processes.
The interaction of these particles with the dense chromospheric plasma produces charged pions ($\pi^\pm$) that decay into a transient neutrino flux extending into the $\mathcal{O}(\text{MeV--GeV})$ range \cite{deWasseige:2016}.
The resulting spectrum can significantly exceed the energies of the steady-state solar neutrino flux and therefore modifies the location of RSFP resonances inside the Sun.
Figure~\ref{fig:resonance_condition} illustrates the energy $E_{\text{res}}$ at which resonance occurs as a function of solar radius, using standard LMA oscillation parameters with $\Delta m^2 \approx 7.5 \times 10^{-5}$ eV$^2$.
For $E \sim 10$ MeV, the resonance lies in the radiative core, in agreement with the conventional picture for $^8$B neutrinos.
When the energy is increased to $E_\nu \sim 1$ GeV, the condition is instead satisfied in the vicinity of the tachocline ($r \approx 0.71 R_\odot$) and within the convective zone.
In other words, the high-energy tail of the flare spectrum effectively shifts the RSFP resonance into the region where the toroidal magnetic field is believed to be strongest.

Adopting the Borexino upper bound on the neutrino magnetic moment, $\mu_\nu = 2.8 \times 10^{-11} \mu_B$, and a representative peak field $B_{\text{max}} = 50$ kG in the tachocline, the corresponding RSFP transition can induce a substantial helicity inversion.
In particular, the polarization vector may evolve from an initially left-handed state with $S_\parallel = -1$ to values close to $S_\parallel \approx 0.04$ near the resonance.
As the neutrino beam propagates further into the convective zone, turbulent fields drive partial depolarization and tend to restore $S_\parallel$ toward more negative values, with the evolution eventually stabilizing around $S_\parallel \approx -0.94$ for the benchmark parameters used in our numerical examples.
\subsection{Unified Helicity Evolution}

To obtain quantitative transition probabilities, we integrate the generalized Bloch equation of Eq.~\eqref{eq:bloch_vector} from the solar interior out to the surface.
The evolution naturally splits into two stages. First, the neutrino undergoes coherent RSFP in the tachocline region, where the large-scale toroidal field dominates.
Then they are submitted to the turbulent magnetic fluctuations and a declining large-scale field jointly drive decoherence and mild depolarization.
In the tachocline, the coherent helicity track approximately follows the effective mixing angle in matter.
Writing the longitudinal polarization in terms of the matter angle as
\begin{equation}
S_\parallel \propto \cos(2\theta_m) =
- \frac{\Omega_3}{\sqrt{\Omega_1^2 + \Omega_3^2}},
\end{equation}
Two limiting regimes are particularly informative:
\begin{itemize}
\item \textbf{Non-adiabatic regime (small $\mu_\nu$).} For sufficiently weak magnetic moments, the adiabaticity parameter is small and $P_{LZ} \to 1$.
In this case the neutrino essentially jumps between eigenstates at resonance and returns to its initial configuration, $S_\parallel \to -1$, producing a nearly symmetric profile.
\item \textbf{Adiabatic regime (large $\mu_\nu$).} For magnetic moments close to the Borexino limit, the transition becomes adiabatic, $P_{LZ} \to 0$.
The polarization vector then follows the adiabatic eigenstate across the resonance, allowing for a significant flip in spin–flavor alignment.
\end{itemize}

Once the neutrino reaches the convective zone ($r \gtrsim 0.73 R_\odot$), turbulent magnetic fluctuations ($\delta B$) act as an open quantum system dissipator.
This induces a steady, exponential decoherence, mildly depolarizing the emerging state, accompanied by high-frequency spatial jitter characteristic of magnetohydrodynamic (MHD) turbulence.
\begin{figure}
    \centering
    \includegraphics[width=0.8\linewidth]{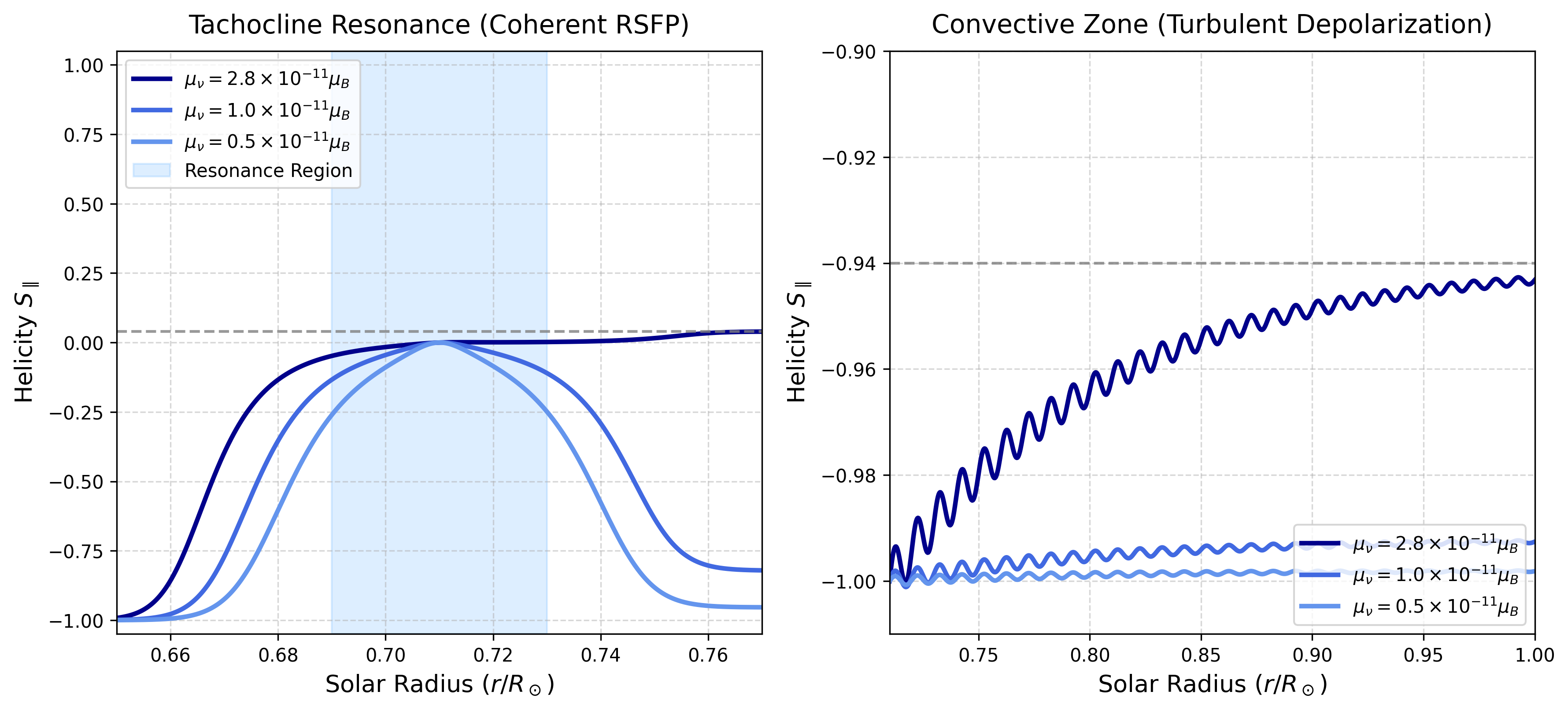}
   \caption{Exact numerical integration of the helicity evolution ($S_\parallel$) for a 1 GeV solar flare neutrino traversing the outer solar layers, evaluated for three neutrino magnetic moment $\mu_\nu$ values.
\textit{Left:} Coherent Resonant Spin-Flavor Precession (RSFP) within the solar tachocline.
In the shaded resonance region, the transition's adiabaticity explicitly scales with $\mu_\nu^2$.
For magnetic moments near the Borexino upper limit ($2.8 \times 10^{-11} \mu_B$), the transition is highly adiabatic, permanently altering the spin-flavor state ($S_\parallel \approx 0$).
Conversely, weaker magnetic moments induce non-adiabatic jumps, forcing the helicity to return to its initial pure left-handed state ($S_\parallel \to -1$).
\textit{Right:} The subsequent evolution of the neutrino state through the solar convective zone.
Interaction with turbulent magnetic field fluctuations acts as an open quantum system dissipator, inducing exponential decoherence accompanied by high-frequency magnetohydrodynamic (MHD) spatial jitter.
The magnitude of this turbulent depolarization scales with the magnetic moment, causing a mild but measurable shift toward a mixed state for larger values of $\mu_\nu$.}
\label{fig:helicity_evolution_exact}
\end{figure}
\begin{figure}
    \centering
    \includegraphics[width=0.8\linewidth]{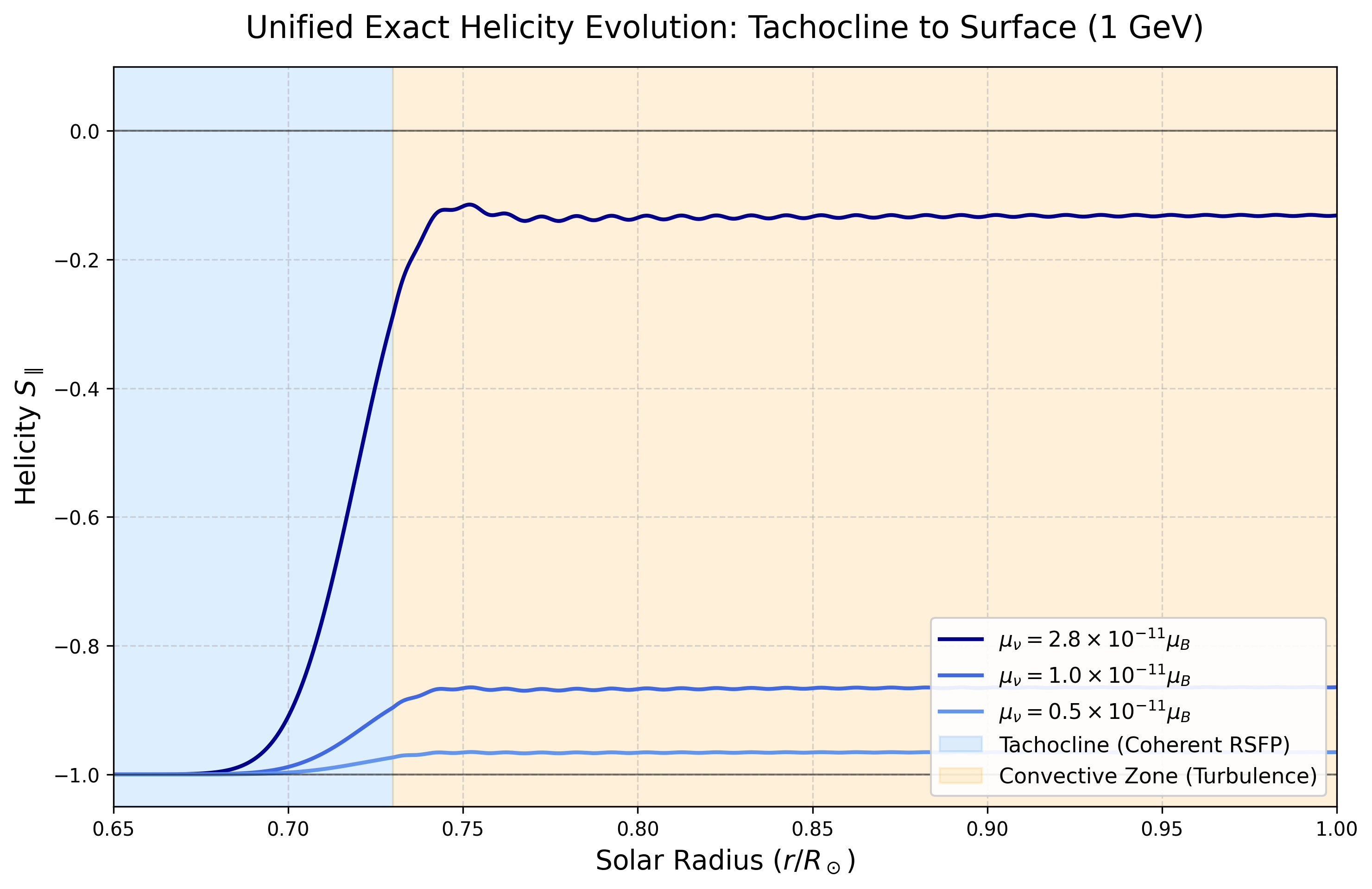}
    \caption{Exact numerical integration of the unified helicity evolution ($S_\parallel$) for a 1 GeV solar flare neutrino propagating from the solar interior to the surface using three neutrino magnetic moment $\mu_\nu$  values.
In the inner region ($r \lesssim 0.73 R_\odot$, shaded blue), the neutrino undergoes coherent Resonant Spin-Flavor Precession (RSFP) driven by the tachocline's localized magnetic field.
The efficiency of this transition is highly sensitive to the magnetic moment: near the Borexino upper limit ($\mu_\nu = 2.8 \times 10^{-11} \mu_B$), the transition is highly adiabatic, inducing a macroscopic, permanent shift in the spin-flavor state.
For weaker magnetic moments, the transition becomes increasingly non-adiabatic, and the neutrino emerges from the tachocline much closer to its initial pure left-handed state ($S_\parallel = -1$).
Upon entering the convective zone ($r > 0.73 R_\odot$, shaded orange), the emergent state is subjected to turbulent depolarization.
This open quantum system environment induces mild exponential decoherence and high-frequency spatial jitter due to magnetohydrodynamic (MHD) fluctuations, establishing the terminal helicity state before the neutrino exits the Sun.}
    \label{fig:unified_helicity_exact_ode}
\end{figure}
\section{Supernova Neutrinos: Envelope Conversion}
\label{sec:supernova_application}

Core-collapse supernovae offer a complementary environment to the Sun, with much higher densities and magnetic fields.
A typical explosion releases a thermal neutrino burst with characteristic energies $E \sim 10$ MeV over a timescale of order 10 s.
If active Dirac neutrinos were to convert efficiently into sterile states deep inside the core, the associated energy loss will be strongly in conflict with the SN1987A signal, leading to bounds on $\mu_\nu$ \cite{Raffelt:2001kv}.
\subsection{Thermal Neutrinos (10 MeV)}

\begin{figure}[ht]
\centering
\includegraphics[width=0.8\linewidth]{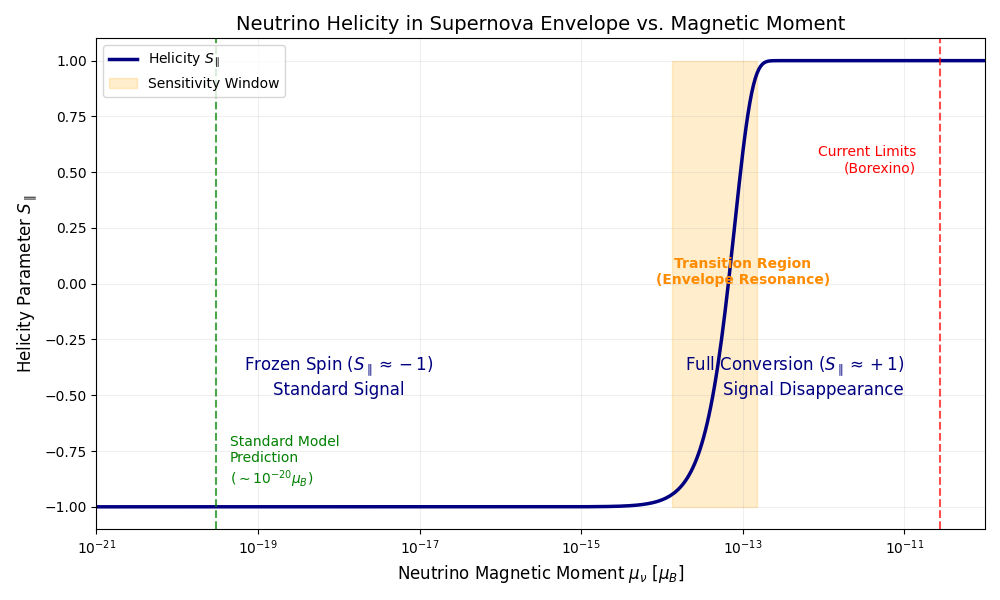}
\caption{Dependence of the final emergent helicity parameter $S_{\parallel}$ on the neutrino magnetic moment $\mu_\nu$ for a 10 MeV thermal neutrino undergoing RSFP in the supernova envelope.}
\label{fig:supernova_envelope}
\end{figure}

For the thermal component with $E \sim 10$ MeV, the resonance density is in the range
\begin{equation}
\rho_{\text{res}} \sim 10^3\text{--}10^4~\text{g/cm}^3,
\end{equation}
corresponding to regions in the outer stellar shells at radii $R \gtrsim 10^3$ km \cite{Woosley:2002zz, Sukhbold:2015wba, Kippenhahn:2012, Dessart:2010ze}. Here the resonance condition is governed by
$\Delta m^2_{\rm atm}$ and the mixing angle $\theta_{13}$, placing
the resonance in the outer stellar envelope where the magnetic fields are
sufficiently strong to ensure $\gamma\gg 1$, in contrast to the standard
solar MSW resonance which involves $\Delta m^2_\odot$ and $\theta_{12}$
at much lower densities.
Since these layers lie well outside the neutrinosphere, the cooling process is not affected by the active-to-sterile conversion in this region.
Magnetic fields in the envelope can reach values as large as $B \sim 10^6$–$10^8$ G \cite{Heger:2004qp, Spruit:2001tb, Gil:2024oxx, Braithwaite:2004kw}, depending on the progenitor and on the magnetic-field amplification mechanism.
For magnetic moments $\mu_\nu \sim 10^{-12} \mu_B$, such fields are sufficient to render the RSFP transition highly adiabatic, with an adiabaticity parameter $\gamma \gg 1$.
In this regime, the helicity state of the neutrino follows the adiabatic eigenstate across resonance and undergoes an almost complete inversion, as illustrated in Fig.~\ref{fig:supernova_envelope}.

From the observational point of view, such an inversion has markedly different consequences for Dirac and Majorana neutrinos.
In the Dirac case, a flip from left-handed active to right-handed sterile states leads to a suppression of the detectable flux at Earth.
For Majorana neutrinos, by contrast, the right-handed state is an active antineutrino, preserving the total event rate but modifying its flavor and helicity composition.
\subsection{High-Energy Neutrinos (1 GeV)}

\begin{figure}[ht]
\centering
\includegraphics[width=0.8\linewidth]{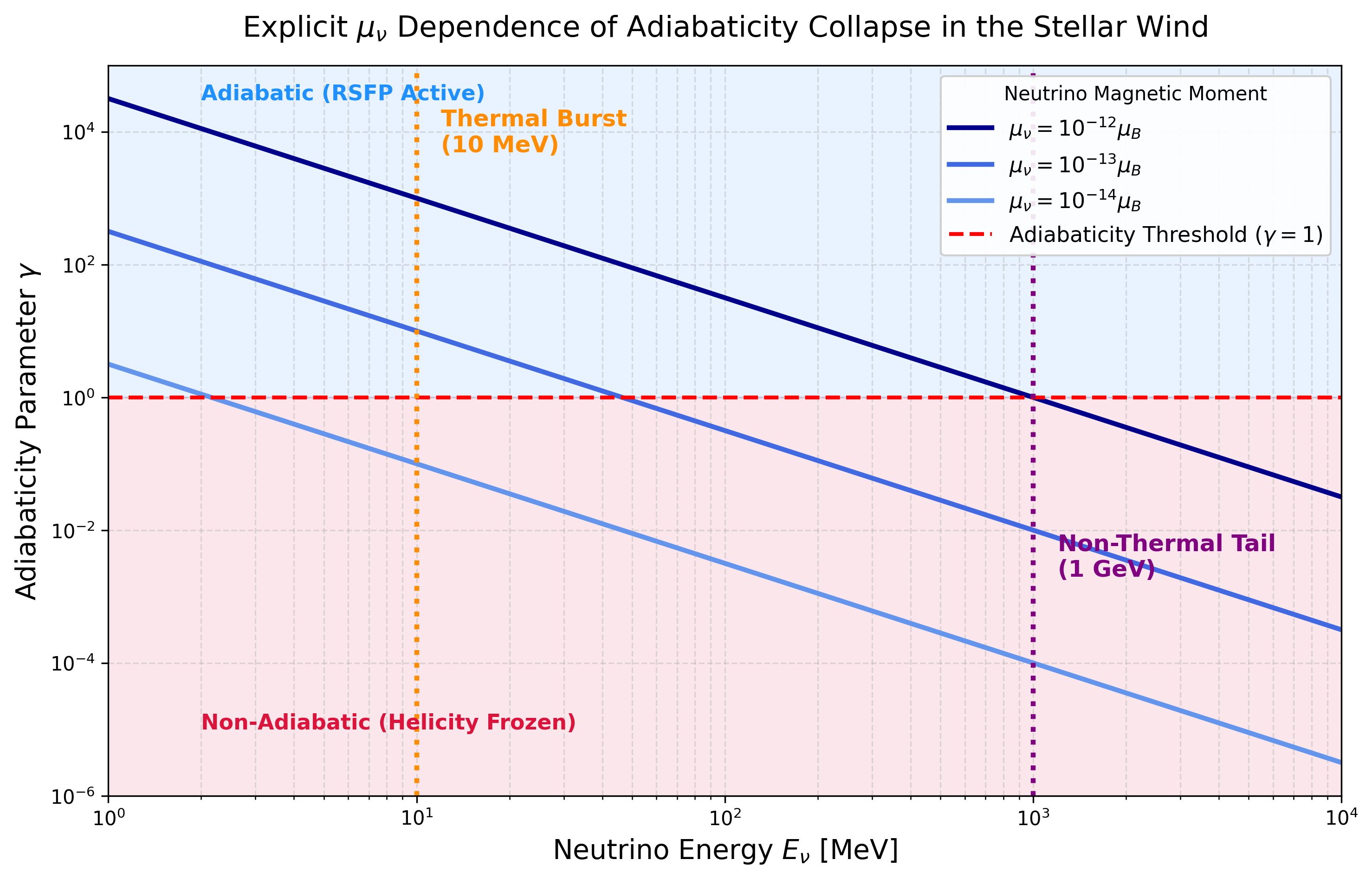}
\caption{Adiabaticity parameter $\gamma$ at the RSFP resonance radius as a function of neutrino energy $E_\nu$.
The $r^{-3}$ decline of the magnetic field in the stellar wind leads to an adiabaticity collapse for the non-thermal 1 GeV tail.}
\label{fig:adiabaticity_collapse}
\end{figure}

Core-collapse supernovae can also produce a non-thermal high-energy neutrino component with $E \sim 1$ GeV.
This component arises when the ejecta interact with the circumstellar medium (CSM) or in the presence of choked jets, leading to pion production and subsequent decay \cite{Murase:2018, Wen:2024}.
The resulting spectrum forms a high-energy tail on top of the thermal burst and typically develops at large radii, where the matter density and magnetic field have both declined.
For neutrinos in this GeV range, the  resonance condition is satisfied at much lower densities,
\begin{equation}
\rho_{\text{res}} \sim 10~\text{g/cm}^3,
\end{equation}
which pushes the resonance far out into the dilute stellar wind.
In this region, the large-scale magnetic field is expected to follow an approximate dipole-like falloff,
\begin{equation}
B(r) \propto r^{-3},
\end{equation}
so that the field strength at the resonance point is strongly suppressed compared to its value in the envelope \cite{Jackson:1998}.
As a consequence, the adiabaticity parameter for GeV neutrinos,
\begin{equation}
\gamma_{\text{GeV}} \propto \frac{\mu_\nu^2 B^2}{|\nabla V_{\text{eff}}|},
\end{equation}
collapses towards zero over a broad range of reasonable parameters.
This behavior is summarized in Fig.~\ref{fig:adiabaticity_collapse}, which shows the adiabaticity parameter $\gamma$ evaluated at the resonance radius as a function of the neutrino energy $E_\nu$.
As the energy increases from MeV to GeV scales, the shift of the resonance into the stellar wind, combined with the steep $r^{-3}$ field falloff, drives the transition from an adiabatic to a strongly non-adiabatic regime.
The helicity parameter for the GeV tail is therefore expected to remain close to its initial value,
\begin{equation}
S_{\parallel}^{\text{GeV}} \approx -1,
\end{equation}
so that RSFP effects on the high-energy component are negligible.

\subsection{Flux Normalization and a Ratio Observable}

A persistent difficulty in supernova neutrino astronomy is the degeneracy between the absolute flux at Earth, the distance to the source $d$, and the total explosion energy.
These uncertainties affect the interpretation of event rates in a given detector and can obscure subtle effects such as those induced by a non-zero $\mu_\nu$.
The survival of the high-energy GeV tail under RSFP suggests a way to mitigate these degeneracies.
We introduce an experimental ratio observable $R$ defined as
\begin{equation}
R = \frac{N_{\text{therm}}^{\text{CE}\nu\text{NS}}(10~\text{MeV})}{N_{\text{CC}}^{\text{GeV}}(1~\text{GeV})},
\end{equation}
where $N_{\text{therm}}^{\text{CE}\nu\text{NS}}$ denotes the number of CE$\nu$NS events induced by the thermal neutrino burst, and $N_{\text{CC}}^{\text{GeV}}$ is the number of charged-current events from the GeV tail.
Both quantities scale as $1/d^2$ with the source distance, so that this dependence cancels in the ratio.
Variations in the overall explosion energy that affect both components in a similar way are also largely removed.
If RSFP in the envelope converts a fraction of the thermal Dirac neutrinos into sterile states, the numerator in $R$ is reduced relative to the Standard Model expectation and the denominator remains unchanged due to the adiabaticity collapse in the GeV tail.
A measured value $R \ll R_{\text{SM}}$ would then point towards Dirac neutrinos with a magnetic moment large enough to induce significant spin–flavor conversion in the envelope.
In contrast, for Majorana neutrinos the total CE$\nu$NS rate is preserved, and $R$ should remain close to its Standard Model value, up to distortions in the recoil spectrum associated with changes in helicity and flavor composition.
In this way, the high-energy non-thermal tail is like an in situ standard candle for the total neutrino luminosity, allowing one to isolate magnetic moment effects on the thermal component with reduced sensitivity to supernova modeling uncertainties.
\section{Quantitative Event Rate Expectations}
\label{sec:event_rates}

The viability of the GeV-scale RSFP scenario hinges on the existence of a sufficiently large neutrino flux and on the capabilities of upcoming detectors.
In this section we provide order-of-magnitude estimates for the relevant event rates and under which conditions the Dirac–Majorana asymmetries described above can become observable are discussed.
\subsection{Solar Flare Neutrino Flux}

In large solar flares, magnetic reconnection accelerates protons and heavier ions that subsequently interact with the chromosphere.
These interactions produce charged pions, whose decays generate a non-thermal neutrino flux typically following an approximate power-law spectrum, $\Phi(E_\nu) \propto E_\nu^{-2}$ or $E_\nu^{-3}$, extending up to GeV energies \cite{deWasseige:2016, deWasseige:2017}.
The flux is not uniform; the differential fluence follows a power-law distribution dictated by the acceleration profile of the initial protons in the magnetic reconnection region.
As a function of energy, it takes the form \cite{deWasseige:2016, deWasseige:2017}:$$\frac{d\Phi}{dE} = \Phi_0 \left( \frac{E}{E_0} \right)^{-\delta}$$
where the spectral index $\delta$ is generally modeled between 2.0 and 3.0.
$E_0$ is the reference energy \cite{deWasseige:2017}.
If we integrate that spectrum over a typical large flare duration, the simulated benchmark fluences at Earth are:
\begin{itemize}
    \item High-Energy Bin ($ > 100$ MeV to a few GeV): Roughly 220 to 780 $\text{cm}^{-2}$.
\item Low-Energy Bin (10 MeV to 100 MeV): Roughly 400 to 770 $\text{cm}^{-2}$.
\end{itemize}
 If the flare only accelerates protons up to 1 GeV, the lower end of that fluence range is hit.
If the acceleration scales up to 5 GeV, pion production increases significantly and the upper bound of roughly 780 $\text{cm}^{-2}$ \cite{deWasseige:2017} is reached.
These hadronic acceleration models confirm that a fluence of $\sim 500 \text{ cm}^{-2}$ extending into the GeV regime is a realistic benchmark \cite{deWasseige:2017}.
Although the overall fluence is modest compared to that of a core-collapse supernova, the close proximity of the Sun and the possibility of temporal coincidence with electromagnetic observations make solar flares an attractive target.
Several studies indicate that, for sufficiently bright events, a combination of large-volume neutrino detectors and temporal gating with gamma-ray observations could allow the detection of tens of neutrino events in the GeV range from an X-class flare at Earth \cite{deWasseige:2017}.
In such a scenario, the RSFP-induced helicity pattern discussed in Sec.~\ref{sec:solar_application} would induce a measurable difference between Dirac and Majorana neutrinos in channels as neutrino–electron scattering and CE$\nu$NS, provided that the integrated statistics are high enough.
Dedicated searches for neutrinos from solar flares have already been performed by IceCube and Super-Kamiokande \cite{IceCube:SolarFlares2021, SuperK:SolarFlares2022}, providing upper limits on the GeV-scale flare fluence.
IceCube's resulting upper limits successfully constrained the theoretical parameter space of expected neutrino fluxes from hadronic acceleration models (which typically predict fluences of $\mathcal{O}(10^2 - 10^3) \text{ cm}^{-2}$ at Earth for large flares) \cite{IceCube:SolarFlares2021}.
In a comprehensive search for solar flare neutrinos across solar cycles 23 and 24, the Super-Kamiokande collaboration analyzed data from massive flares, including the record-breaking X28.0 flare from November 4, 2003 and they obtained a 90\% confidence level (C.L.) upper limit on the solar flare neutrino fluence at Earth of $\Phi < 1.1 \times 10^6 \text{ cm}^{-2}$ \cite{SuperK:SolarFlares2022}.
\subsection{Supernova High-Energy Non-Thermal Tail}

For core-collapse supernovae, the high-energy non-thermal tail is generated when the ejecta interact with the circumstellar medium or when internal shocks in choked jets accelerate hadrons that subsequently produce pions \cite{Murase:2018}.
The ensuing pion-decay neutrinos can extend into the GeV–TeV range and may be detectable by current and future high-energy neutrino telescopes.
To determine the experimental potential to observe the effects mentioned in this paper, one needs to know the fluence of the thermal burst supernova neutrinos ($E_{\nu} \approx 10$ MeV) and the fluence of the non-thermal supernova neutrinos ($E_{\nu} \approx 1$ GeV).
For the thermal burst neutrino, one expects a Fluence at Earth (assuming that the burst was at 10 kpc distance from Earth): $\mathcal{O}(10^{11} \text{ to } 10^{12}) \text{ cm}^{-2}$ \cite{Scholberg:2012,Lang:2016}.
Because the fluence is absolutely massive, it easily overcomes the small interaction cross-section at 10 MeV.
This flux will yield roughly 8,000 events in a 50-kiloton water Cherenkov detector like Super-Kamiokande, and tens of thousands of CE$\nu$NS events in next-generation dark matter detectors.
For the High-energy non-thermal neutrino tail, one expects a Fluence at Earth at 10 kpc distance from Earth: $\mathcal{O}(10^{3} \text{ to } 10^{5}) \text{ cm}^{-2}$ \cite{Murase:2018,Wen:2024}. While the raw particle count (fluence) is roughly eight orders of magnitude smaller than the thermal burst, the neutrino interaction cross-section increases linearly with energy.
Therefore, this GeV-scale fluence is still sufficient to produce $\sim 10 \text{ to } 100$ events in gigaton/megaton-scale observatories like IceCube or Hyper-Kamiokande.
\subsection{Statistical Significance}

To assess the statistical requirements for distinguishing Dirac and Majorana scenarios, we consider the total number of detected events
\begin{equation}
N = \int \Phi(E_\nu)\, \sigma(E_\nu)\, N_{\text{targ}}\, \epsilon(E_\nu)\, dE_\nu,
\end{equation}
where $\Phi(E_\nu)$ is the neutrino flux, $\sigma(E_\nu)$ the relevant cross section, $N_{\text{targ}}$ the number of targets in the detector, and $\epsilon(E_\nu)$ the detection efficiency.
Assuming Poisson statistics, an asymmetry $\mathcal{A}$ between Dirac and Majorana event rates becomes statistically significant when
\begin{equation}
N \gtrsim \frac{1}{\mathcal{A}^2}.
\end{equation}
The requirement $N \gtrsim 1/\mathcal{A}^2$ should be understood as a minimal statistical benchmark;
in practice, the control of detector systematics will be equally important.
The recent CONUS+ analysis \cite{DeRomeri:2025CONUSPlus} illustrates that sub-$10\%$ uncertainties in CE$\nu$NS measurements are already within reach at reactor energies, suggesting that analogous performance may be attainable for a nearby galactic supernova in next-generation detectors.
Facilities such as Hyper-Kamiokande \cite{Abe:2018uyc, Abe:2021cMw} and IceCube-Gen2 \cite{Aartsen:2015yva, Aartsen:2020fgd, Abbasi:2021jcq} are expected to collect event samples of this order, or larger, for a galactic core-collapse supernova or a very bright solar flare.
In addition, coordinated multi-messenger observations can enhance sensitivity by providing external triggers and narrowing the time window for the search.
\subsection{Probing New Parameter Space}

The sensitivity to the neutrino magnetic moment ultimately depends on the minimal helicity flip probability that can be resolved.
For a given asymmetry threshold $\mathcal{A}_{\text{min}}$, one can estimate the corresponding requirement on the Landau–Zener parameter and on the combination $\mu_\nu B$ characterizing RSFP.
Resolving an asymmetry at the level $\mathcal{A} \gtrsim 10\%$ typically requires a spin-flip probability $P_{\text{flip}} \gtrsim 0.05$.
Using the Landau–Zener relation, this translates into an adiabaticity parameter $\gamma \gtrsim 0.033$.
As a consequence, one obtains a rough detectability condition
\begin{equation}
\mu_\nu B \gtrsim 5.0 \times 10^{-8}~\mu_B \cdot \text{G}.
\end{equation}
For tachocline fields of order $B \approx 50$ kG, this corresponds to sensitivities around
\begin{equation}
\mu_\nu \gtrsim 10^{-12} \mu_B.
\end{equation}

\begin{figure}[ht]
\centering
\includegraphics[width=0.8\linewidth]{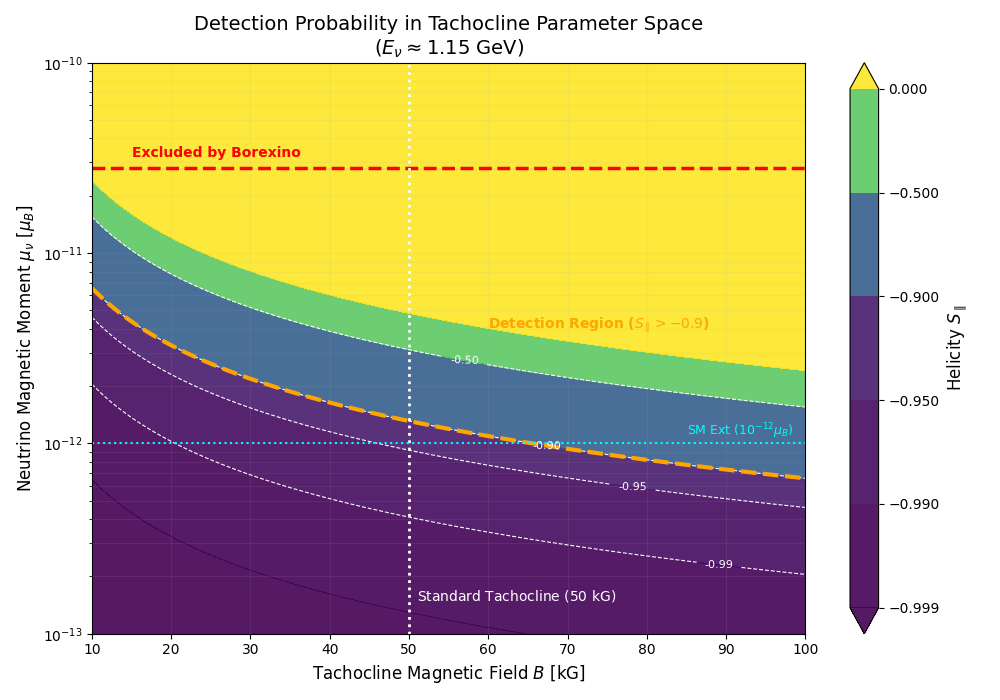}
\caption{Contour plot of the neutrino helicity parameter $S_{\parallel}$ for 1 GeV solar flare neutrinos in the solar tachocline, as a function of magnetic field strength and neutrino magnetic moment.
The dashed line marks the Borexino upper limit on $\mu_\nu$.}
\label{fig:munu_B}
\end{figure}

These estimates are illustrated in Fig.~\ref{fig:munu_B}, which shows contours of the helicity parameter $S_\parallel$ for 1 GeV solar flare neutrinos as a function of $\mu_\nu$ and $B$, with the Borexino upper limit indicated for reference.
The region where RSFP-induced asymmetries become sizable is within the parameter space region below current direct bounds, suggesting that GeV-scale astrophysical transients could provide a complementary probe of neutrino magnetic moments.
Laboratory constraints on neutrino magnetic moments from CE$\nu$NS and reactor experiments are continuously improving, as shown by the recent CONUS+ results \cite{DeRomeri:2025CONUSPlus}.
While such measurements probe $\mu_\nu$ at low energies and short baselines, the GeV-scale astrophysical strategy discussed here is sensitive to the same parameter in a very different regime, with an effective reach down to $\mu_\nu \gtrsim 10^{-12}\,\mu_B$ for tachocline fields of order $B \sim 50$ kG.

\section{Scattering Cross Sections and Asymmetries}
\label{sec:scattering}

The helicity evolution at GeV energies discussed in the previous sections manifests itself in measurable differences between Dirac and Majorana neutrinos in terrestrial scattering experiments.
In this section we summarize the relevant cross sections for neutrino–electron scattering and CE$\nu$NS, and we show how the final helicity parameter $S_\parallel$ feeds into observable asymmetries.
\subsection{Neutrino–Electron Scattering}

For ultra-relativistic solar neutrinos with $E_\nu \gg m_\nu$, the differential cross sections for scattering on an electron with longitudinal spin component $S_\parallel$ can be written as \cite{Rosen:1982pj, Kayser:1982br, Bernabeu:2000hf}
\begin{align}
\frac{d\sigma_D}{d\Omega} &\propto (1 - S_{\parallel}) \left[ Y(\theta) + Z(\theta) \right], \\
\frac{d\sigma_M}{d\Omega} &\propto 2 \left[ Y(\theta) - S_{\parallel} Z(\theta) \right].
\end{align}
Here $Y(\theta)$ and $Z(\theta)$ denote kinematic functions that depend on the scattering angle $\theta$ and on the vector and axial couplings,
\begin{align}
Y(\theta) &= (g_V^2 + g_A^2) \left[ 1 + \frac{(E_e + E_\nu \cos \theta)^2}{s} \right], \\
Z(\theta) &= 2 g_V g_A \left[ 1 - \frac{(E_e + E_\nu \cos \theta)^2}{s} \right],
\end{align}
with $E_e$ the initial electron energy and $s$ the center-of-mass invariant.
The key point is that the Dirac rate carries an overall factor $(1 - S_\parallel)$, whereas in the Majorana case the dependence on $S_\parallel$ appears only in the interference term proportional to $Z(\theta)$.
To quantify the difference between Dirac and Majorana cases, we define a normalized asymmetry
\begin{equation}
\mathcal{A} = \frac{\sigma_M - \sigma_D}{\sigma_M + \sigma_D},
\end{equation}
where $\sigma_{D,M}$ are the cross sections integrated over the relevant angular range.
In the absence of spin precession, $S_\parallel = -1$ and the Dirac and Majorana cross sections coincide, so that $\mathcal{A} = 0$.
\subsection{CE$\nu$NS and the Supernova Ratio Observable}

For CE$\nu$NS, the differential cross section with respect to the nuclear recoil energy $T$ can be written schematically as \cite{Freedman:1973yd, Kopeliovich:1974qv, Drukier:1983gj, Akimov:2017ade, Plestid:2020ods}
\begin{align}
\frac{d\sigma_D}{dT} &\propto (1 - S_{\parallel})\, Q_W^2\, F^2(T), \\
\frac{d\sigma_M}{dT} &\propto 2\, Q_W^2\, F^2(T),
\end{align}
where $Q_W$ is the weak nuclear charge and $F(T)$ is the nuclear form factor.
Since $Q_W$ is the same for neutrinos and antineutrinos, the Majorana scattering rate is insensitive to the helicity state, while the Dirac rate is directly suppressed when a fraction of the flux is converted into sterile states.
The ratio observable $R$ introduced in Sec.~\ref{sec:supernova_application},
\begin{equation}
R = \frac{N_{\text{therm}}^{\text{CE}\nu\text{NS}}}{N_{\text{CC}}^{\text{GeV}}},
\end{equation}
can therefore be interpreted as a diagnostic of RSFP-induced deficits in the thermal Dirac flux.
A measured value $R \ll R_{\text{SM}}$ would point to Dirac neutrinos with a magnetic moment large enough to drive significant conversion of the thermal component into sterile states, whereas $R \approx R_{\text{SM}}$ with a hardened recoil spectrum \cite{Gelmini:1999mi} would be more naturally associated with Majorana neutrinos.
Table~\ref{tab:unified_summary} summarizes the qualitative expectations across the different astrophysical environments considered in this work, for representative values of the magnetic moment and field strength.
\begin{table}[ht]
\centering
\caption{Summary of theoretical RSFP outcomes in the environments considered here.
The emergent helicity $S_\parallel^{\text{final}}$ is evaluated for a benchmark magnetic moment $\mu_\nu \sim 10^{-12} \mu_B$.}
\label{tab:unified_summary}
\renewcommand{\arraystretch}{1.3}
\begin{tabular}{@{}lcccc@{}}
\toprule
\textbf{Astrophysical source} & \textbf{Energy} & \textbf{Resonance location} & \textbf{Adiabaticity ($\gamma$)}  & \textbf{Asymmetry ($\mathcal{A}$)} \\ \midrule
Standard solar ($^8$B) & 10 MeV & Radiative core ($r < 0.2 R_\odot$) & $\gamma \ll 1$  & $\sim 0\%$ \\
Solar flare (transient) & 1 GeV & Tachocline ($r \approx 0.71 R_\odot$) & $\gamma \gg 1$  & $ \sim 27.7\%$  (CE$\nu$NS) \\
Supernova thermal burst & 10 MeV & Stellar envelope ($R > 10^3$ km) & $\gamma \gg 1$  & $\sim 18\%$ (CE$\nu$NS) \\
Supernova non-thermal tail & 1 
GeV & Dilute stellar wind & $\gamma \to 0$  & Standard candle \\ \bottomrule
\end{tabular}
\end{table}

\subsection{Numerical Summary for 1 GeV Solar Flare Neutrinos}

Using the terminal helicity states $S_\parallel^{\text{final}}$ obtained from the numerical integration of the Bloch equations, one can estimate the size of the Dirac–Majorana asymmetries for different values of the magnetic moment.
Table~\ref{tab:unified_asymmetry} reports representative results for 1 GeV solar flare neutrinos traversing the tachocline and the convective zone.
\begin{table}[ht]
\centering
\caption{Representative numerical results for 1 GeV solar flare neutrinos, including the final helicity parameter $S_\parallel^{\text{final}}$, the resulting neutrino–electron scattering asymmetry $\mathcal{A}$, and the CE$\nu$NS Dirac deficit, for different values of $\mu_\nu$.}
\label{tab:unified_asymmetry}
\renewcommand{\arraystretch}{1.3}
\begin{tabular}{@{}lccc@{}}
\toprule
\textbf{Neutrino magnetic moment ($\mu_\nu$)} & $\mathbf{S_\parallel^{\text{final}}}$ & \textbf{$e$–$\nu$ asymmetry ($\mathcal{A}$)} & \textbf{CE$\nu$NS Dirac deficit} \\ \midrule
$2.8 \times 10^{-11} \mu_B$ (Borexino limit) & $-0.1313$ & $14.07\%$ & $43.43\%$ \\
$1.0 \times 10^{-11} \mu_B$ & $-0.8644$ & $2.20\%$ & $6.78\%$ \\
$0.5 \times 10^{-11} \mu_B$ & $-0.9653$ & $0.56\%$ & $1.73\%$ \\ \bottomrule
\end{tabular}
\end{table}

Even for magnetic moments below the Borexino limit, the table shows that a non-zero asymmetry persists, albeit at a reduced level.
This suggests that, given sufficient statistics and careful control of systematics, solar flare observations at GeV energies could provide an independent handle on $\mu_\nu$ and on the Dirac–Majorana character of the neutrino.
The event-rate and asymmetry benchmarks discussed here are consistent with the level of precision already demonstrated in CE$\nu$NS measurements at CONUS+ \cite{DeRomeri:2025CONUSPlus} and with the sensitivities inferred from IceCube and Super-Kamiokande searches for neutrinos from solar flares \cite{IceCube:SolarFlares2021, SuperK:SolarFlares2022}.
Complementary strategies to distinguish Dirac from Majorana neutrinos at CE$\nu$NS facilities have been explored in Refs.~\cite{Hati:2022CEvNS}.
\subsection{Numerical Summary for Supernova Neutrinos}
For a standard core-collapse supernova, the $10\text{ MeV}$ thermal burst neutrinos resonate within the outer stellar envelope.
Because the remnant magnetic fields here are still sufficiently strong ($B \sim 10^6 - 10^8\text{ G}$), the transition is highly adiabatic ($\gamma \gg 1$).
This drives a macroscopic, nearly complete helicity inversion, shifting the neutrinos from pure left-handed states to right-handed states ($S_\parallel \to +1$).
When these neutrinos reach terrestrial dark matter detectors (which utilize flavor-blind Coherent Elastic Neutrino-Nucleus Scattering, or CE$\nu$NS), one has two cases:
\begin{itemize}
    \item Majorana Neutrinos: Convert into active right-handed antineutrinos, which retain their full scattering cross-section.
The signal remains at the Standard Model baseline.
    \item Dirac Neutrinos: Convert into sterile right-handed states that completely evade weak detection.
This results in an effective signal reduction (or deficit) in the Dirac channel.
\end{itemize}
A core-collapse supernova emits roughly equal total energy (luminosity) into all six neutrino species: $\nu_e, \bar{\nu}_e, \nu_\mu, \bar{\nu}_\mu, \nu_\tau,$ and $\bar{\nu}_\tau$. Because the RSFP resonance depends on the matter potential of the stellar envelope ($V_e$), only the electron flavors hit the resonance.
If all flavors had the exact same energy spectrum, losing only the electron neutrinos would result in exactly a 33.3\% deficit in the total CE$\nu$NS event rate.
In reality, the deficit will likely be slightly lower than 33\% (usually modeled around 25\% to 30\%) due to the difference of energy between flavor neutrinos.
The $\nu_x$ flavors ($\mu$ and $\tau$) do not interact via Charged Current in the core, so they decouple deeper inside the star where it is hotter.
Their average energy is roughly $\langle E_{\nu_x} \rangle \approx 15$ MeV.
The $\nu_e$ and $\bar{\nu}_e$ remain trapped longer and decouple further out where it is cooler.
Their average energy is roughly $\langle E_{\nu_e} \rangle \approx 11$ MeV \cite{Lang:2016,Mirizzi:2016,Keil:2003}.
As the surviving heavy flavors are significantly hotter, they inherently produce more scattering events in the detector per particle than the cooler electron flavors do and as a consequence, the total event rate drops by only $\sim 25 - 30\%$ in place of strictly $33.3\%$.
If we assume a realistic total CE$\nu$NS deficit of 30\%, the total asymmetry becomes:$$\mathcal{A} \approx \textbf{17.6\%}$$.
\section{Conclusion}
\label{sec:conclusion}

In this work, we have demonstrated that the energy-dependent spatial shift of the Resonant Spin-Flavor Precession (RSFP) condition provides a powerful, multi-environment model for determining the fundamental nature of the neutrino.
Standard searches utilizing thermal ($\sim 10$ MeV) neutrinos are severely limited: solar $^8$B neutrinos resonate in the deep core where RSFP is strongly constrained by Borexino data, and core-based conversion of Dirac supernova neutrinos is definitively excluded by the cooling duration bounds of SN1987A.
By studying the $\sim 1$ GeV energy regime—accessible via transient solar flares and the non-thermal tail of core-collapse supernovae—and using the scaling of the resonance density with neutrino energy ($\rho_{\text{res}} \propto E^{-1}$) one obtains the following results:
\begin{enumerate}
    \item \textbf{Solar Flare Transients:} the 1 GeV resonance shifts outward into the solar tachocline and convective zones.
Here, the strong toroidal magnetic fields ($B \sim 50$ kG) induce highly efficient, adiabatic spin-flavor conversion.
This generates a macroscopic normalized asymmetry of $\mathcal{A} \approx 14\%$ between Dirac and Majorana differential cross-sections in terrestrial neutrino-electron scattering.
\item \textbf{Supernova High-Energy Tail:} to avoid the cooling supernovae constraints, the RSFP is moved from the core supernovae to the stellar envelope. The 1 GeV resonance shifts into the dilute stellar wind where magnetic fields collapse, suppressing adiabaticity and "freezing" the helicity state.
This non-thermal flux remains unaltered by RSFP, allowing it to be used as an \textit{in situ} standard candle.
Normalizing the flavor-blind thermal CE$\nu$NS signal against this high-energy tail cancels underlying astrophysical uncertainties and reveals a massive signal deficit ($\sim 30\%$) uniquely characteristic of Dirac sterile conversion.
\end{enumerate}

We have established that the required event rates to resolve these asymmetries are well within the observational reach of next-generation observatories.
By utilizing multi-messenger triggers (e.g., HAWC gamma-ray alerts \cite{Abeysekara:2013tza, Abeysekara:2018qxs, Linden:2020bkz} or SNEWS \cite{Antonioli:2004zb, Scholberg:2012}) to temporally gate the analysis and suppress steady-state atmospheric backgrounds, detectors such as Hyper-Kamiokande and IceCube-Gen2 require only a modest number of events to achieve statistical significance. 

Our approach transforms high-energy astrophysical transients into precision laboratories. The observation of these predicted scattering asymmetries would help to settle the Dirac-Majorana debate.
Conversely, a null observation under this temporally-gated model would improve current upper bounds on the neutrino magnetic moment by more than two orders of magnitude, probing down to $\mu_\nu \sim 10^{-14} \mu_B$.

\begin{acknowledgments}
We acknowledge financial support from SECIHTI and SNII (M\'exico).
\end{acknowledgments}
\appendix
\section{Derivation of the Generalized Bloch Equation}

To clarify the transition from the Lindblad master equation to the vector form used in our numerical analysis, we provide the explicit derivation here. We begin with the parametrization of the density matrix for a two-level flavor-spin system in terms of the polarization vector $\vec{P}(t)$:
\begin{equation}
\rho(t) = \frac{1}{2} \left[ P_0(t) \mathbb{I} + \vec{P}(t) \cdot \vec{\sigma} \right]
\end{equation}
where $\mathbb{I}$ is the identity matrix and $\vec{\sigma}$ represents the triplet of Pauli matrices. Substituting this into the master equation:
\begin{equation}
i \frac{d\rho}{dt} = [H_{eff}, \rho] - i \mathcal{D}[\rho]
\end{equation}

\subsection{Coherent Evolution}
The effective Hamiltonian $H_{eff}$ can be decomposed in the Pauli basis as $H_{eff} = \frac{1}{2} \vec{\Omega} \cdot \vec{\sigma}$, where $\vec{\Omega}$ is the effective precession vector. The commutator term becomes:
\begin{equation}
[H_{eff}, \rho] = \frac{1}{4} [(\vec{\Omega} \cdot \vec{\sigma}), (\vec{P} \cdot \vec{\sigma})] = \frac{1}{4} \Omega_j P_k [\sigma_j, \sigma_k]
\end{equation}
Using the commutation identity $[\sigma_j, \sigma_k] = 2i\epsilon_{jkl}\sigma_l$, we obtain:
\begin{equation}
[H_{eff}, \rho] = \frac{i}{2} \epsilon_{jkl} \Omega_j P_k \sigma_l = \frac{i}{2} (\vec{\Omega} \times \vec{P}) \cdot \vec{\sigma}
\end{equation}

\subsection{Dissipative Term and Vector Form}
The dissipator $\mathcal{D}[\rho]$ encodes environmental effects such as collisions and MHD turbulence. In the relaxation time approximation, the environment induces decoherence in the off-diagonal elements of the density matrix. This process is  described by the damping of the polarization components transverse to the precession axis:
\begin{equation}
-i \mathcal{D}[\rho] \rightarrow -\frac{1}{2} (\Gamma \vec{P}_{\perp} \cdot \vec{\sigma})
\end{equation}
where $\Gamma$ is the decoherence rate. Combining the coherent and dissipative parts and equating the coefficients of $\vec{\sigma}$ on both sides, we arrive at the generalized Bloch equation :
\begin{equation}
\frac{d\vec{P}}{dt} = \vec{\Omega} \times \vec{P} - \Gamma \vec{P}_{\perp}
\end{equation}
This confirms that the neutrino state evolves through a combination of coherent Larmor precession and environmental depolarization.
%

\end{document}